\documentclass[aps,prb,twocolumn,showpacs,amsmath,amssymb,unsortedaddress]{revtex4-1}

\usepackage[version=3]{mhchem} 
\usepackage{graphicx}
\usepackage{dcolumn}
\usepackage{bm}
\usepackage{url} 
\usepackage[colorlinks=true,linkcolor=blue,citecolor=blue,urlcolor=blue]{hyperref} 
\usepackage{multirow}

\usepackage{color}

\newcolumntype{.}{D{.}{.}{-1}}

\newcommand*{\bR}{{\bf R}}
\newcommand*{\br}{{\bf r}}

\begin{document}

\title{Towards a systematic assessment of errors in diffusion Monte Carlo calculations of semiconductors: case study of zinc selenide and zinc oxide}

\author{Jaehyung Yu$^1$}
\author{Lucas K. Wagner$^2$}
\author{Elif Ertekin$^{1,3}$}
\affiliation{$^1$Department of Mechanical Science \& Engineering, 1206 W Green Street, University of Illinois at Urbana-Champaign, Urbana IL 61801}
\affiliation{$^2$Department of Physics, University of Illinois at Urbana-Champaign, Urbana IL 61801}
\affiliation{$^3$International Institute for Carbon Neutral Energy Research (WPI-I$^2$CNER), Kyushu University, 744 Moto-oka, Nishi-ku, Fukuoka 819-0395, Japan}

\email[e-mail: ]{ertekin@illinois.edu}

\begin{abstract}
  The fixed node diffusion Monte Carlo (DMC) method has attracted interest in recent years as a way to calculate properties of solid materials with high accuracy. 
  However, the framework for the calculation of properties such as total energies, atomization energies, and excited state energies is not yet fully established.  
  Several outstanding questions remain as to the effect of pseudopotentials, the magnitude of the fixed node error, and the size of supercell finite size effects.  
  Here, we consider in detail the semiconductors ZnSe and ZnO and carry out systematic studies to assess the magnitude of the energy differences arising from controlled and uncontrolled approximations in DMC.  
  The former include time step errors and supercell finite size effects for ground and optically excited states, and the latter include pseudopotentials, the pseudopotential localization approximation, and the fixed node approximation. 
  We find that for these compounds, the errors can be controlled to good precision using modern computational resources, and that quantum Monte Carlo calculations using Dirac-Fock pseudopotentials can offer good estimates of both cohesive energy and the gap of these systems.
We do however observe differences in calculated optical gaps that arise when different pseudopotentials are used. 
\end{abstract}


\maketitle

\section{Introduction} 

Increasing demand for quantitative predictions of material properties, coupled with growing complexity of materials of current research interest, promotes the development of high-accuracy methods from first principles.  
Recently the quantum Monte Carlo (QMC) method has emerged as a viable approach to quantitative calculations of the properties of real materials with chemical identity\cite{ErtekinMgO,Santana2015,ZhengWagner2015,cerium,perovskite,tio2polymorph,aluminum,ErtekinWinkler,hydrogen}.  
QMC methods comprise a suite of tools for calculation of material properties {\it via} stochastic solution of the many-particle Schr{\"o}dinger equation\cite{Foulkes2001,Petruzielo2012,Kolorenc2011,NeedsTowler2010}.  
Because of their direct treatment of electron correlation, QMC methods are amongst the most accurate available and have a history of ground breaking, benchmark calculations\cite{Grossman2001}.  

However, despite the successes of QMC thus far, an open question still remains as to the practical accuracy of the technique.  
In practice uncertainties arise from several approximations such as the use of pseudopotentials and the pseudopotential localization error\cite{Shulenburger2015,Casula06}, finite size effects\cite{KZK2008,KentHoodWilliamson99,Chiesa97,drummondfinitesize}, and the fixed node error that is present in fixed node diffusion Monte Carlo\cite{AndersonFN}.  
At present, limitations in our understanding of the coupled effect of these uncertainties make practical usage of the technique challenging and hinder progress. 
The purpose of this manuscript is to present a systematic assessment of the uncertainties from these competing factors.  
As a case study, we have considered the semiconductors zinc selenide and zinc oxide in detail, and we quantify the size of uncertainties coming from the factors listed above. 

\begin{table*}
    \begin{tabular}{c|c|p{0.3\textwidth}|p{0.3\textwidth}}
    {\bf Approximation } & {\bf Controlled}  & {\bf Description } & {\bf Assessment method}  \\ 
    \hline
    \hline
    Pseudopotential & No & Replace the core electrons with an effective potential & Vary within the space of reasonable potentials \\
    \hline
    Localization & No & Approximate the diffusion Monte Carlo projector in the presence of a nonlocal potential & Vary trial function and projector approximations \\
    \hline
    Nodal & No & Fix the zeros of the trial wave function when performing diffusion Monte Carlo & Vary the trial wave function, apply variational theorem. \\
    \hline
    Timestep & Yes & Approximate diffusion Monte Carlo projector & Reduce time step until quantities are converged. \\
    \hline
    Finite size & Yes & Finite size cells with periodic boundary conditions  & Increase the supercell size and average over twisted boundary conditions \\
    \hline
  \end{tabular}
  \caption{Sources of error in fixed node diffusion Monte Carlo that are considered in this work.  The column ``controlled" means that the error can be reduced to zero with a polynomially-scaling amount of computer time. }
  \label{table:approximations}
\end{table*}

Variational Monte Carlo (VMC) and diffusion Monte Carlo (DMC) are two  common approaches within the QMC suite of methods \cite{Foulkes2001,Petruzielo2012,Kolorenc2011,NeedsTowler2010}.  
In variational Monte Carlo, an explicitly correlated form of the many-body wave function $\Psi$ is used ({\it e.g.} Slater-Jastrow) and the expectation value $\langle \Psi | \hat{H} | \Psi \rangle / \langle \Psi | \Psi \rangle$ is evaluated stochastically ($\hat{H}$ is the many-body Hamiltonian).  
The parameters of the wave function can be optimized by either energy or variance minimization.  
In diffusion Monte Carlo (DMC), the wave function is described by a finite number of electron configurations  (walkers).  
The time-dependent Schr{\"o}dinger equation is mapped onto a diffusion equation in imaginary time, and the walkers are propagated according to the dynamics of the diffusion equation.  
This approach uses Green's function methods to project out the ground state distribution of walkers from the trial wave function (with some complexities to be discussed later).  
Several recent demonstrations of the capability of QMC include the calculation of optical transitions and thermal ionization levels for $F$-center defects in MgO\cite{ErtekinMgO}, intrinsic and extrinsic defects in zinc oxide\cite{Santana2015}, metal to insulator transition in VO$_2$\cite{ZhengWagner2015},  the volume collapse in cerium \cite{cerium}, perovskite and post-perovskite MgSiO$_3$ in the earth's lower mantle \cite{perovskite}, and several others \cite{tio2polymorph,aluminum,ErtekinWinkler,hydrogen}.

Despite these promising studies, a clear set of ``best practices'' is not yet well-established for the for the calculation of material properties such as total energies, band gaps, and defect properties within the fixed node DMC (DMC) framework.  
Understanding the relative size of uncertainties arising from competing factors like pseudopotentials, fixed node errors, and finite size effects will be an important aspect of making quantum Monte Carlo a standard computational tool.  
A recent comprehensive analysis of QMC applied to solids has demonstrated the viability of the technique, focusing on assessing the QMC prediction of lattice constants and equilibrium volumes across a extensive spectrum of materials (including metals, semiconductors, and insulators)\cite{Shulenburger2013}.  
In this work, we take a complementary approach focusing instead on two materials in detail to assess competing sources of error in the calculation of total energies, atomization energies, and band gaps. 
While some aspects of this analysis may be known by experts, there are very few existing works that systematically assess these factors.  This manuscript is intended to serve as a reference for those who wish to carry out these calculations.  
Also, even amongst practitioners there remains considerable discussion related to decoupling uncertainties arising from these effects.

We chose zinc selenide and zinc oxide because they are both reasonably well understood, compound, wide band gap semiconductors from the II-VI family.  
The presence of zinc enables assessment of errors in the presence of localized states (the 10 $3d$ electrons of Zn). 
The comparison of the oxide to the selenide gives a chance to consider distinct effects arising from different chemistries since oxygen is more electronegative than selenium (ZnO is more ionic).  
As a semiconductor, ZnO has several desirable properties including good transparency, high electron mobility, and strong room temperature luminescence. 
It is used as a transparent conducting electrode in liquid crystal displays and photovoltaic cells, and in electronics for thin-film transistors and light-emitting diodes.  
A large outstanding challenge is the elusiveness of obtaining p-type ZnO in the natively n-type material. 
For ZnSe, technological uses include II-VI light-emitting diodes (blue emission)~\cite{Eason1995}, infrared laser gain mediums (when doped with Cr)~\cite{DeLoach1996}, infrared optical materials exhibiting a wide transmission wavelength range, and scintillators (when doped with Te)~\cite{Schotanus1992}.  

\section{Method}

A brief sketch of the quantum Monte Carlo methods used here follows. More details can be found in Refs~[\onlinecite{Foulkes2001,Petruzielo2012,Kolorenc2011,NeedsTowler2010}]. We concentrate on explaining the methods well enough so that the approximations are clear; there are many details of implementation that affect the efficiency dramatically but do not affect the accuracy.
Variational Monte Carlo is a direct implementation of the variational method for correlated wave functions.
In this work, we use the Slater-Jastrow wave function,
\begin{equation}
  \Psi_{SJ}(\bR)=\text{Det}[\phi_k(\br_i^\uparrow)]\text{Det}[\phi_k(\br_i^\downarrow)] \exp[\sum_{i,j,\alpha} u(r_{i\alpha}, r_{j\alpha}, r_{ij}) ],
  \label{eqn:slater_jastrow}
\end{equation}
where ${\phi_k}$ are one-particle orbitals obtained from a DFT calculation, ($i,j$) refer to electron indices, $\alpha$ is a nuclear/ion index, $\bR$ is the many-body electron coordinate $(\br_1,\br_2,\ldots,\br_N)$, and $u$ is the same as the one in Ref. [\onlinecite{lkw2007}]. 
The parameters in the $u$ function are optimized either using variance or energy minimization.
The function $u$ can take many forms, most common are two-body Jastrow functions which explicitly include electron-electron interactions and three-body Jastrow functions which explicitly include electron-electron-nucleus interactions.  

To obtain higher accuracy, we use fixed node diffusion Monte Carlo (DMC). 
In this method, starting with a trial function $\Psi_T$ (in this work $\Psi_T = \Psi_{SJ}$), a projection to the ground state $|\Phi_0\rangle$ is performed:
\begin{equation}
  \langle \bR | \Phi_0 \rangle = \lim_{\tau\rightarrow \infty} 
  \int{\langle\bR|\exp[-\hat{H}\tau] | \bR'\rangle \langle \bR'| \Psi_T \rangle } d\bR'
\end{equation}
where $\hat{H}$ is the Hamiltonian. 
In principle, this integral can be done by Monte Carlo integration.
In practice, there are two major impediments to the projection.
First, the matrix element $\langle\bR|\exp[-\hat{H}\tau] | \bR'\rangle$ is only known in the limit of small $\tau$. 
This is solved by using the identity $\exp[-\hat{H}\tau]=\left(\exp[-\hat{H}\delta t]\right)^{\tau/\delta t}$ and inserting many resolution of identity operators to increase the dimensionality of the integral. $\delta t$ is the time step in diffusion Monte Carlo.
Second, the Hamiltonian matrix element includes a sign for fermions, which causes an exponentially decreasing signal to noise ratio with system size. 
This is the sign problem, which can be resolved in general only with an approximation. 
We use the common choice of the fixed node approximation, in which $\Phi_{FN}(\bR)=0$ wherever $\Psi_T(\bR)=0$. 
The resulting method obtains a variational upper bound to the ground state energy, with equality when the nodes of the trial wave function are equal to the nodes of the ground state wave function. 

While in principle DMC is a ground state method, the fixed node constraint allows approximate access to excited states. 
Under some conditions, a variational upper bound to the excited state energy can be obtained\cite{FoulkesGaps}, and in practice calculated excitation gaps can be quite accurate, even for challenging highly correlated materials\cite{Kolorenc2011}.
We perform excited state calculations by promoting an electron from the valence band maximum to the conduction band minimum in the Slater determinant in Eq.~(\ref{eqn:slater_jastrow}) and proceeding as outlined for the ground state.

In Table~\ref{table:approximations}, we present the major approximations present in the DMC technique, which we classify as either {\it controlled} or {\it uncontrolled} based on whether the uncertainty can be reduced to zero with a polynomially-scaling computer time or not. 
We will examine each of these one by one for the case of the semiconductors ZnSe and ZnO.  In all cases for the work presented here, trial wave functions $\Psi_T$ for the QMC simulations were generated using the DFT framework as implemented within the CRYSTAL code\cite{CRYSTAL14}.  
For the QMC calculations, we use the QWalk\cite{QWalk} package. 

In our CRYSTAL calculations, we systematically optimize the localized basis sets using line optimization. 
The Gaussian exponents and weights are varied, and we select the basis set that gives the minimum energy.  
The basis sets used here are available in supplemental material~\cite{supplemental}. 
When this optimization scheme is used, our final DMC energies are not particularly sensitive to the basis set (the variation of the DMC energies are within error bars). 
Additionally, for all DMC simulations we use a sufficient number of walkers so that population bias is reduced to less than the stochastic error bars of our results. 
In our simulations, we use the experimental lattice constants for the supercells of the materials in question. 

\section{Results and Discussion}

\subsection{Uncontrolled Approximations}

\subsubsection{Pseudopotentials}

\begin{table}[h] 
  \caption{Total energies (eV) of isolated atoms Zn, Se, and O according to DFT (PBE0) and DMC for both Trail--Needs (TN) and Burkatski-Filippi-Dolg (BFD) pseudopotentials.} 
   \label{table:iso-atoms}
     \begin{tabular}{ c | c | c | c  } 
       {\bf Pseudo } & {\bf Atom } & {\bf DFT(PBE0)} & {\bf DMC} \\  \hline \hline 
 \multirow{3}{*}{TN} & Zn  & -1773.59 & -1744.48(3) \\ \cline{2-4}
                     & Se  & -253.25 &  -252.96(1)  \\ \cline{2-4}
                     & O   & -429.63 &  -431.25(1)  \\ \hline \hline 
   \multirow{3}{*}{BFD} & Zn  & -6179.75 &  -6178.57(5) \\ \cline{2-4}
                        & Se  & -253.01 &  -252.49(1)  \\ \cline{2-4}
                        & O   & -430.79	 &  -432.47(1)  \\ 
   \end{tabular}
\end{table}

\begin{table}[h]
  \caption{Atomization energy of zincblende ZnSe (top) and wurtzite ZnO (bottom) with experimental lattice constants according to DFT (PBE0) and DMC for both Trail--Needs (TN) and Burkatski--Filippi--Dolg (BFD) pseudopotentials, in comparison to experiment. } 
   \label{table:znse}
     \begin{tabular}{ c | c | .  }
       {\bf Pseudo } & {\bf Method} & \multicolumn{1}{c}{{\bf Atomization }}  \\ 
                              &  & \multicolumn{1}{c}{{\bf Energy (eV/fu)}}  \\  \hline \hline 
    TN  & DFT(PBE0) &  5.71  \\ \hline
     BFD & DFT(PBE0) &  5.13 \\  \hline \hline
       TN  & DMC &  5.54(9) \\ \hline   
       BFD  & DMC &   5.68(8) \\ \hline  \hline    
    Experiment~\cite{MatTherm} & -- & 5.511  \\   
   \end{tabular}
\vspace{0.5cm}
   \label{table:zno}
     \begin{tabular}{ c | c | .  }
       {\bf Pseudopotential } & {\bf Method} & \multicolumn{1}{c}{{\bf Atomization }}  \\ 
                              &  & \multicolumn{1}{c}{{\bf Energy (eV/fu)}}  \\  \hline \hline 
    TN  & DFT(PBE0) & 7.83  \\ \hline
     BFD & DFT(PBE0) &  8.63 \\  \hline \hline
       TN  & DMC &  7.61(8)  \\ \hline     
       BFD  & DMC &  7.67(8) \\ \hline  \hline     
    Experiment~\cite{MatTherm} & -- & 7.59  \\   
      \end{tabular}
\end{table}

The optimal choice of pseudopotentials for quantum Monte Carlo calculations has emerged as an important question in recent years~\cite{Shulenburger2015,Mitas3}. 
In the pseudopotential approximation, each electron is classified as either a core or valence electron, and the former are assumed largely inert, while the latter are significantly perturbed by bonding.  
The use of pseudopotentials eliminates the need to directly include the core electrons in the simulation and makes the calculation tractable, but it is an approximation and in reality a well-defined boundary between ``core'' and ``valence"  does not exist.  

We use relativistic Hartree-Fock ({\it i.e.} Dirac-Fock) pseudopotentials. There are several examples in the literature that show they are well-suited for diffusion Monte Carlo simulations of solids \cite{RaschMitas,ZhengWagner2015,cerium,KolorencMitasPRL,Shulenburger2015}. 
We consider Dirac-Fock pseudopotentials from two sources:  the Burkatski-Filippi-Dolg~\cite{BFD1,BFD2} and Trail-Needs~\cite{TN1,TN2} sets.  
These do not include core polarization or spin-orbit coupling. 
We focus our discussion on the zinc pseudopotential in particular, since the transition metal element with a full $3d^{10}$ set of electrons is the more challenging and interesting case than O and Se by comparison. 
The Ne-core Burkatski-Filippi-Dolg Zn pseudopotential has 20 electrons in valence while the Ar-core Trail-Needs Zn pseudopotential leaves 12 electrons in valence. 
The comparison of the large and small core pseudopotential for Zn allows us to assess the extent to which the deeper semicore (Zn $3s$ and $3p$) levels are perturbed by the bonding. 
There may be other differences between the two pseudopotentials as well, since they are constructed by different authors and not exactly in the same way.

In Table~\ref{table:iso-atoms} we show the DMC and DFT-PBE0 total energies of the isolated Zn, Se, and O atoms for both the TN and BFD pseudopotential. 
As a test of pseudopotential accuracy, we show the atomization energy calculated within DMC in Table \ref{table:zno} for ZnSe and ZnO. 
For completeness we also show the atomization energies according to DFT-PBE0, but note that these do not necessarily give indications of pseudopotential accuracy, since the DFT prediction of the atomization energy itself may be wrong.  
The DMC total energies are obtained by finite size supercell extrapolation and the T-moves scheme (both described in the next sections).
The DFT results in Table \ref{table:zno} are obtained using the PBE0 approximation to the exchange correlation functional, and for the most part both the BFD and the TN pseudopotential give quite reasonable results within PBE0.  
For ZnSe, in comparison to the experimental value of 5.51 eV/fu (formula unit)~\cite{MatTherm}, the TN pseudopotential slightly overestimates at 5.71 eV/fu while the BFD pseudopotential slightly underestimates at 5.13 eV/fu.  
In both cases, DMC improves the description of the atomization energy: we obtained 5.54(9) eV/fu for TN and 5.68(8) eV/fu for BFD. 
We observed similar trends in Table \ref{table:zno} for the atomization energy of ZnO (7.59 eV/fu in experiment). 
Here, for both pseudopotentials DFT-PBE0 results overbind the solid relative to isolated atoms: the degree of overbinding is 0.24 eV/fu for TN pseudopotentials and quite large (1.04 eV/fu) for BFD pseudopotentials.  
Once again, DMC improves the description substantially for both cases: TN gives 7.61(8) eV/fu and BFD gives 7.67(8) eV/fu.  
In all cases, the DMC results yield atomization energies that are within 0.1 eV/atom of the experimental value. 

For these materials, it appears that the large-core TN pseudopotential obtains similar results as the small-core BFD pseudopotential when using QMC, although we do not {\it a priori} expect this to hold true in general.
Note that from Table \ref{table:zno}, it is not true for DFT (PBE0).  
As we will discuss later, it is also not the case for the DMC calculation of the optically excited state.
Atomic cores are more perturbed for highly ionic semiconductors, which may necessitate use of the small core pseudopotentials. 
Since ZnO is more ionic, for the remainder of the article we assess both TN and BFD potentials for that material, and only the TN potential for ZnSe. 

\subsubsection{Localization error}

\begin{figure}[h]
\includegraphics[width=7.5cm]{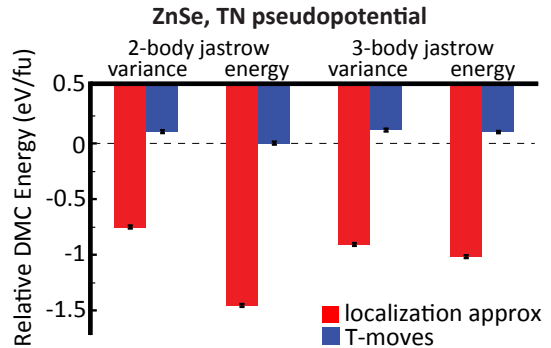}
\caption {Relative energy per formula unit ZnSe according to DMC, obtained by varying Jastrow factor and optimization method, both within the localization approximation and with T-moves. In the localization approximation, total energy differences can vary by as much as 0.75 eV for different Jastrow forms and optimization methods.  By contrast, however, the variation is within error bars amongst all combinations when T-moves is used. The use of T-moves reduces localization errors, without substantial cost in the calculation time. Energies are shown relative to the minimum calculated DMC energy when T-moves is used.}
  \label{tmoves}
\end{figure}

\begin{figure}[h]
\includegraphics[width=7.5cm]{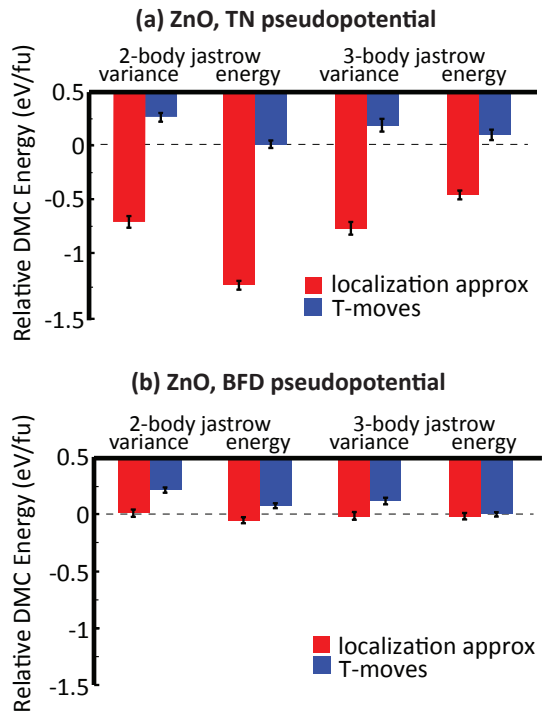}
\caption {Relative energy per formula unit of ZnO according to DMC, obtained by varying Jastrow factor and optimization method, both within the localization approximation and with T-moves.  Within the localization approximation, localization errors are reduced with the small core BFD pseudopotential compared to the large core TN pseudopotential. Energies are shown relative to the minimum calculated DMC energy when T-moves is used.} 
  \label{tmoves-zno}
\end{figure}

Nonlocal pseudopotentials introduce extra terms in the imaginary-time Green's function $\langle R | \exp[-\hat{H}\tau] |R'\rangle$ in DMC. 
These terms would cause a sign problem, which can be removed either using the localization approximation\cite{Mitas91} or the T-moves scheme\cite{Casula06}.  
Both of these approximations introduce an error in addition to the fixed node error that depends on the quality of the trial wave function. 
When the trial wave function approaches the exact one, the localization error approaches zero.

To assess the degree of the localization error, we generated different trial wave functions within VMC by considering (i) two and three body Jastrow functional forms, and (ii) energy and variance optimization of the Jastrow parameters.  
There are four possible combinations, and for all four trial wave functions we evaluated the total energy in DMC within the localization approximation and with T-moves.  
Fig. \ref{tmoves} shows the DMC energy per fu ZnSe using the TN pseudopotential for all four trial wave functions.  The energies computed within the localization approximation (red bars) can vary by up to 0.75 eV/fu from one scheme to the other, which shows a large dependence of the projected out wave function on the different forms of the trial wave function.  
By contrast, the incorporation of T-moves results in more uniform total energies, which now vary within error bars instead (blue bars).  
Additionally, with T-moves the recovery of the variational theorem would allow unambiguous determination of the best choice for the trial wave function, although in this case all four trial wave functions give statistically equivalent results.  
It is notable that the use of T-moves appears to reduce the dependence of the final DMC total energy on the trial wave function.  

In Fig. \ref{tmoves-zno}, we show the DMC relative energy per fu for ZnO for all four trial wave functions, but also compare the TN (Fig. \ref{tmoves-zno}a) and the BFD (Fig. \ref{tmoves-zno}b) pseudopotential.  
The results for the TN pseudopotential in ZnO are similar to those of Fig. \ref{tmoves} for ZnSe: within the localization approximation the DMC total energies can vary by around 0.75 eV/fu whereas they are much more uniform when T-moves is used.  
Interestingly, by contrast, in Fig.~\ref{tmoves-zno}b for the small core BFD pseudopotential, the sensitivity of the total energies on the trial wave function is much smaller (in fact, within statistical error) within the localization approximation. 
In this case both the localization approximation and T-moves appear to give similar results, which is consistent with our expectation that localization errors are smaller when the pseudopotential core is smaller.

For the remainder of the paper, all results presented have been obtained using T-moves and energy optimization to minimize the localization error. 

\subsubsection{Nodes of the trial wave function}

\begin{figure}[h]
\includegraphics[width=7.5cm]{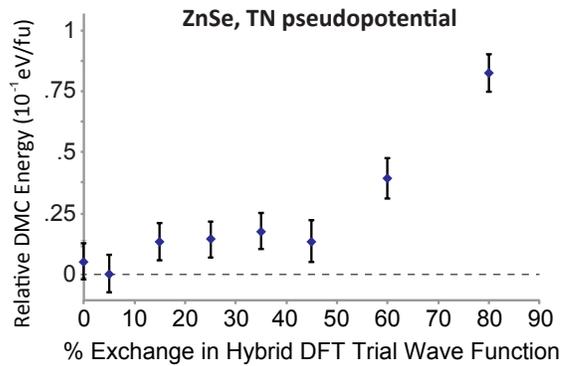}
  \caption {DMC energy for ZnSe as a function of the degree of exchange $\alpha$ used to generate the trial wave function in DFT.  For smaller $\alpha$ from $0\% \le \alpha \le \sim 50\%$, the DMC total energies exhibit only a small sensitivity.  For larger $\alpha$, the DMC energy increases by around 0.1 eV/fu at most, indicating that the nodal structure is not as good. Total energies are shown relative to the minimum calculated energy across all $\alpha$ considered.} 
  \label{nodal}
\end{figure}

\begin{figure}[h]
\includegraphics[width=7.5cm]{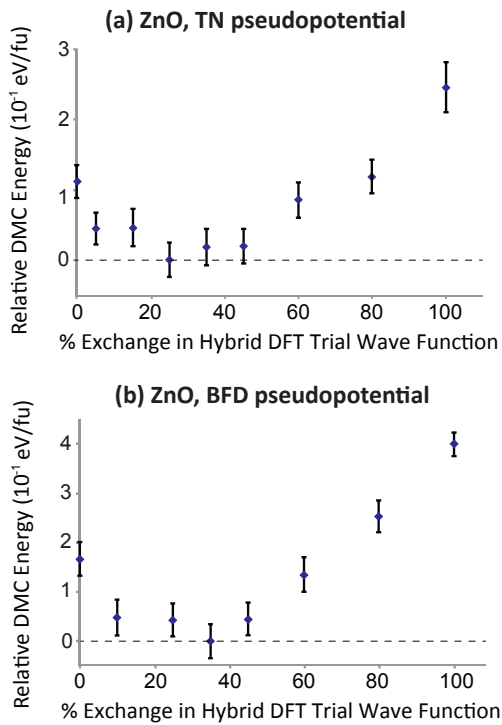}
  \caption {DMC energy for ZnO as a function of the degree of exchange $\alpha$ used to generate the trial wave function in DFT using (a) Trail-Needs and (b) Burkatski-Filippi-Dolg pseudopotentials.  For both cases there appears to be a minimum around $\alpha \approx 20\%-30\%$, indicating a better nodal structure.  Total energies are shown relative to the minimum calculated energy across all $\alpha$ considered.} 
  \label{nodalzno}
\end{figure}

The fixed-node approximation is another source of error within the fixed node DMC framework. 
During a fixed node DMC simulation, the nodes of the trial wave function are held fixed to preserve the antisymmetric nature of the wave function and avoid collapse to the bosonic ground state \cite{AndersonFN}.  
To accomplish this, in practice the DMC walkers are prevented from crossing the nodes during propagation.  
If the pseudopotentials are sufficiently accurate and other controllable sources of error such as finite size effects are addressed adequately, then the accuracy of the DMC approach will in principle be limited by the accuracy of the trial wave function nodes. 
If the trial wave function has the exact nodal structure, then DMC will project out the exact ground state. 
However, if the nodal structure differs from the exact one, then the DMC algorithm will converge to the closest approximation of the ground state subject to the constrained nodes. 
Since the inexact solution has an energy higher than the exact solution, in principle the nodal surface can be optimized by minimizing the total energy (although this is challenging in practice). 
At present, a detailed understanding of the magnitude of the fixed node error is somewhat lacking, particularly for solids, although some recent headway has been made~\cite{Mitas2}.  More is known about the effects of nodal errors within molecular and/or atomic systems~\cite{Mitas2,Mitas1}.  

To consider the sensitivity of the final DMC energy on the nodal structure, Fig.~\ref{nodal} shows the relative energy of ZnSe (eV/fu) obtained from DMC calculations using trial wave functions generated from orbitals from hybrid DFT calculations, with different degrees of exchange mixing $\alpha$ within the PBE1$_x$ framework. 
The exact nodal structure of the trial wave functions is not known, but it is believed to span a reasonable range since the orbitals are generated from theories ranging from DFT-PBE ($\alpha = 0$\%) to something similar to Hartree-Fock ($\alpha=100$\%).
For these two materials, the main effect of the exchange mixing is to modify the relative energies of the $p$ and $d$ orbitals, which affects their hybridization and the resultant Kohn-Sham orbitals that are used in the Slater determinant.
Because DMC is variational, the lowest energy obtained is the closest upper bound to the exact energy and so we minimize with respect to $\alpha$.
This approach of varying the exchange weight is often used to obtain better trial wave functions for DMC \cite{ZhengWagner2015,Mitas4,Wagner2014}; some authors vary the $U$ parameter in DFT+$U$ to achieve a similar effect \cite{PhysRevX.4.031003}.

For the case of ZnSe, from Fig.~\ref{nodal} the sensitivity of the total energy to $\alpha$ is very small: for $0\% \le \alpha < 50\%$ the total energies are within error bars and vary by only around ~0.02 eV/fu.  
This variation is much smaller than the localization errors in the previous section.  
Only for $\alpha > 50\%$ does the DMC energy increase, but never by more than 0.1 eV/fu for the full range of $\alpha$ considered here.  
Overall, this indicates that for ZnSe the sensitivity to the nodal structure of the Slater determinant is small. 
On the other hand for the case of ZnO shown in Fig. \ref{nodalzno} the sensitivity is still small, but for both pseudopotentials there appears to be a minimum somewhere around $\alpha \approx 20\%-30\%$. 
In some other semiconducting systems (MnO~\cite{JoshMnO}) we have also found an optimal $\alpha \approx 25$\% and energy variations around 0.1 eV/fu with varying $\alpha$.  
This is also consistent with other reported findings for transition metal 3d compound FeO \cite{Mitas4}. 
The DMC energies do not vary by more than 0.1 eV/fu  for $0\% \le \alpha < 50\%$, but for $\alpha = 100$\% the energy has increased by around 0.3 to 0.4 eV/fu above the minimum for both pseudopotentials. 
Although the sensitivity of the DMC energy to the nodal structure appears to be larger for ZnO (the more ionic compound) than ZnSe, it is still small in comparison to other effects such as the pseudopotential localization approximation considered above.    

Regarding nodal errors, it will be interesting in the future to establish in detail which solid materials exhibit a greater nodal sensitivity and which do not. 
It will also be interesting to assess whether nodal sensitivity is greater for excited state wave functions for which the nodal surface may be more complex. 
Additionally, we emphasize that it is still an open question how to most effectively further optimize the nodes of the wave function. 

\subsection{Systematically Controllable Approximations} 

We now turn to the controllable approximations, for which the errors can be made small by converging a simulation parameter. 
There are several controllable approximations in QMC, which include the finite DMC time step, one and many-body finite size effects, the quality of the basis set, and the number of configurations.
The basis set and the number of configurations are relatively easy to converge; we consider the more challenging approximations. 
We present here results for the DMC time step error, the many-body finite size effect for the calculation of ground state total energies, and the finite size effect for the calculation of optical excitation energies. 

\subsubsection{DMC time step error}

\begin{figure}[h]
\includegraphics[width=7.5cm]{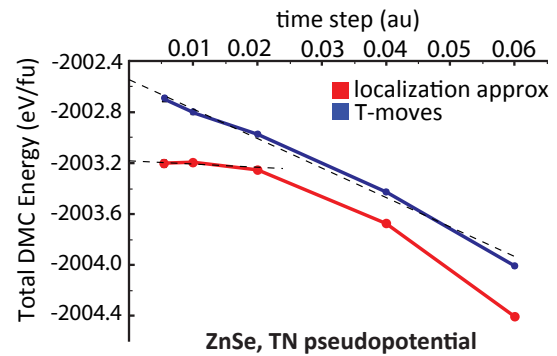}
  \caption {Relative DMC energy of ZnSe calculated from an 8 atom supercell at the $\Gamma$ point, as as a function of time step. As the time step decreases, the acceptance ratio increases and total energy converges towards a fixed value. Stochastic uncertainties are smaller than the symbol size. The linear extrapolation is shown both for the use of T-moves and the localization approximation, demonstrating that the two approximations give different extrapolated values of the energy.}
  \label{timestep}
\end{figure}

In DMC, the time-dependent Schr{\"o}dinger equation is transformed into an imaginary-time diffusion equation that includes a source/sink term.  
A large number of ``walkers", which are the possible configurations of all $N$ electrons distributed into the $3N$ dimensional phase space (each walker constitutes a $3N$ dimensional set of coordinates) according to the trial wave function. 
The walkers are then propagated according to the dynamics of the imaginary time diffusion equation using a Green's function approach. 
The Green's function projector is rigorously only exact for vanishingly small time step, but the propagation of walkers in practice requires a finite time step.  
This finite time step introduces an error in the projected energy \cite{AndersonFN,Umrigar93}. Controlling the time step error is typically straightforward and can be accomplished by performing calculations for a range of time steps and extrapolating the result to the limit of zero time step. 
The tradeoff is that smaller time steps require more total number of steps for sufficient phase space sampling.  
If the time step is sufficiently small, the results will exhibit a linear dependence of the total energy on the time step since it the linear term is the first term in the expansion.

In Fig.~\ref{timestep}, the dependence of the total DMC for ZnSe on the DMC time step is shown for two cases: when the localization approximation is used, and when T-moves is used.  
Both sets of results show the expected trends for the dependence of the DMC energy on the timestep; we show the linear fit to the calculated results in the linear regime.
Note that the extrapolated values of the DMC energy differ from each other for the two methods.  
In the linear regime the localization approximation appears to have smaller time step errors than the T-moves scheme, although we should note that we did not implement the recent version of the algorithm\cite{casula_size-consistent_2010}, which might have smaller time step errors. 
For time steps $< 0.01$ au, the error in the calculated total energy is $< 0.1$ eV/fu, using the algorithms implemented in the QWalk package.
The results for ZnO (not shown) are similar to those shown in Fig.~\ref{timestep} for ZnSe. 

\subsubsection{Supercell finite size effects for ground state energies}

\begin{figure}[h]
\begin{tabular}{l} 
\includegraphics[width=7.5cm]{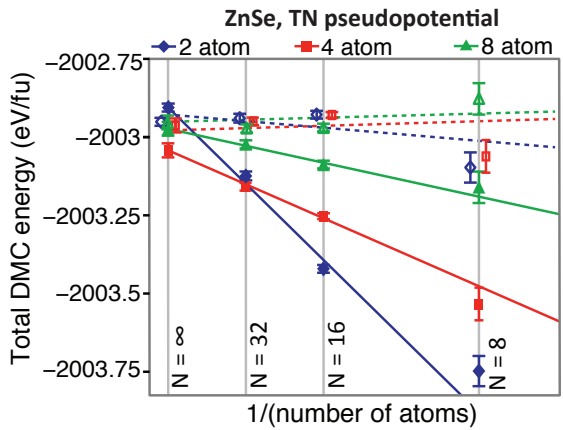} \\
\end{tabular} 
\caption {Extrapolation of total energy of ZnSe to thermodynamic limit, using unit cells of different shapes and sizes. Solid lines indicate the calculated DMC energies while dashed lines include the structure factor correction.}
  \label{ZnSetotalenergy}
\end{figure}

\begin{figure}[h]
\begin{tabular}{l} 
(a) \\
\includegraphics[width=7.5cm]{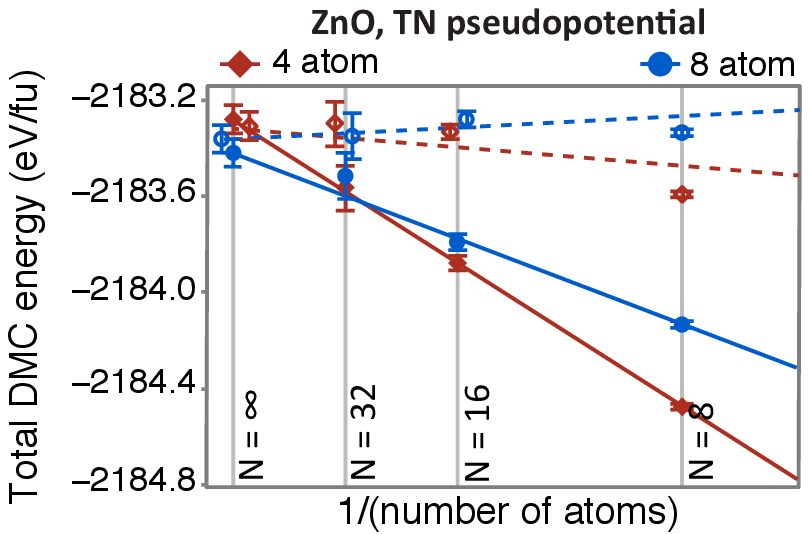} \\
(b)\\
\includegraphics[width=7.5cm]{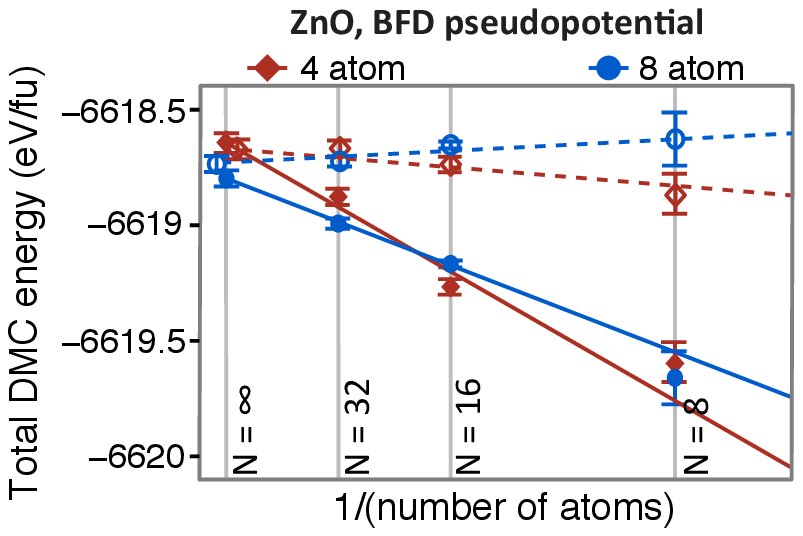}
\end{tabular} 
\caption {Extrapolation of total energy of ZnO with (a) TN, and (b) BFD pseudopotential to thermodynamic limit. Solid lines indicate the calculated DMC energies while dashed lines include the structure factor correction.} 
  \label{ZnOtotalenergy}
\end{figure}

Simulations of solids often invoke periodic boundary conditions applied to a unit cell or a supercell of the solid materials.  
Finite-size effects in QMC can take many forms, in particular (i) insufficient twists included for the wave function boundary conditions (the analog of insufficient $k$-point sampling in single particle theories) and (ii) the many-body finite size effect, which arises from spurious electron-electron image interactions amongst electrons in neighboring cells (this finite size effect is unique to many-particle theories and has no analog in single-particle descriptions). 
Regarding the former, averaging over twists (phases of the wave function) results in a faster convergence to the thermodynamic limit as the supercell size increases.  
 For all of our supercells we have twist-averaged over a $2\times2\times2$ grid. 
 Since the wave functions for these twists are real, this grid allows for extrapolation to larger cells due to the improved computational efficiency.

The many-body finite size effect is of more interest: the fictitious electron-electron image interaction introduces an artificial correlation that has a stabilizing effect on the total energy. 
This correlation depends on both the size and the shape of the supercell and affects the calculated total energy. 
Possible correction schemes have been discussed in the past \cite{KZK2008,KentHoodWilliamson99,Chiesa97,drummondfinitesize}, but a systematic investigation across a variety of materials carrying out extrapolations to large sized supercells is still lacking due to the computational cost of DMC.  
Here, we illustrate the magnitude of the many-body finite size effect, and apply the structure-factor $S(k)$ based correction scheme of Chiesa {\it et al.}\cite{Chiesa97}

Figs.~\ref{ZnSetotalenergy} and~\ref{ZnOtotalenergy} show the results for ZnSe and ZnO  respectively. The solid lines are the uncorrected total energies, while the dashed lines include the structure factor correction. 
In these figures the $x$-axis shows $1/N$, where $N$ is the number of atoms in the supercell, since the leading order correction for the many-body finite size effect scales as $1/N$.
In Fig.~\ref{ZnSetotalenergy} the blue dots correspond to $(2 \times 2 \times 1)$, $(2 \times 2 \times 2)$, and $(2 \times 2 \times 4)$ supercells of the 2-atom zincblende fcc unit cell.  
The red dots correspond to $(2 \times 1 \times 1)$, $(2 \times 2 \times 1)$, and $(2 \times 2 \times 2)$ supercells built from a 4-atom (wurtzite-like) building block, again giving systems of total size 8, 16, and 32 atoms. 
Finally, the green dots correspond to $(1 \times 1 \times 1)$, $(2 \times 1 \times 1)$, and $(2 \times 2 \times 1)$ supercells of the cubic 8 atom conventional zincblende unit cell, also giving rise to supercells of 8, 16, and 32 atoms.  
In total, for ZnSe we have considered 9 supercells, 3 each of 8 atoms, 16 atoms, and 32 atoms.
Similarly in Fig. ~\ref{ZnOtotalenergy} for ZnO the magenta dots correspond to $(2 \times 1 \times 1)$, $(2 \times 2 \times 1)$, and $(2 \times 2 \times 2)$ supercells built from a 4-atom wurtzite building block. 
The blue dots correspond to $(1 \times 1 \times 1)$, $(2 \times 1 \times 1)$, and $(2 \times 2 \times 1)$ supercells of the 8 atom tetragonal unit cell.  
For both cases, the lines show the best fit linear extrapolations of each set of data points to the thermodynamic limit $N \rightarrow \infty$. 
The expected $1/N$ scaling is observed, and all extrapolations yield total ground state energies within 0.15 eV/fu of each other. 

According to Figs.~\ref{ZnSetotalenergy},~\ref{ZnOtotalenergy} the many-body finite size effect is large, but the application of the structure factor correction (dashed lines in Figs.~\ref{ZnSetotalenergy},~\ref{ZnOtotalenergy}) helps substantially. 
For instance for 8 atom supercells of different shapes the total energy per fu may be 0.5 --1 eV lower than the extrapolated limit for both ZnSe and ZnO. 
For both however, most of the $1/N$ dependence is eliminated and the linear fits correspondingly flatten out when the structure factor correction is included. 
In all cases, we find that the use of 16 or 32 atom supercells, together with $S(k)$ correction, are sufficient to resolve energies to within 0.15 eV/fu. 
Further, it is encouraging that all extrapolations, both with and without $S(k)$ correction tends towards similar values within $\approx$ 0.15 eV/fu of each other.

\subsubsection{Supercell finite size effects for optical excitations}

\begin{figure}[h]
\begin{tabular}{l} 
\includegraphics[width=8.5cm]{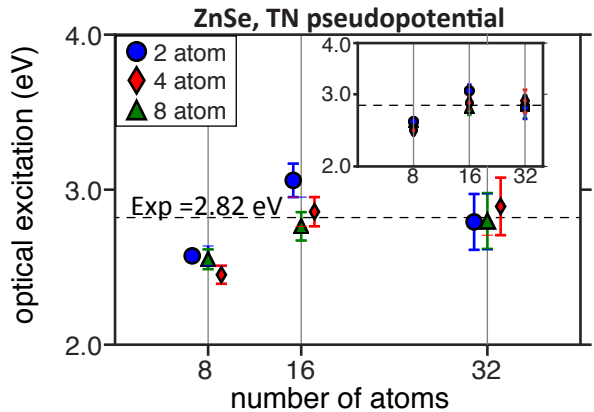} \\    
\end{tabular} 
\caption {Optical excitation energy computed for different supercell sizes and shapes for ZnSe. The excitation energy is obtained as the difference of the $\Gamma$ point ground state and first optically excited state energies.} 
  \label{ZnSeoptenergy}
\end{figure}

\begin{figure}[h]
\begin{tabular}{l} 
(a) \\
\includegraphics[width=8.5cm]{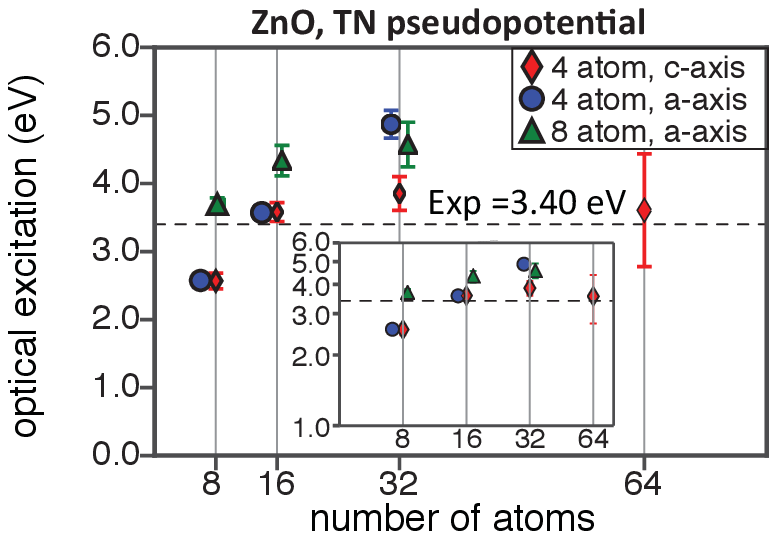} \\    
(b) \\ 
\includegraphics[width=8.5cm]{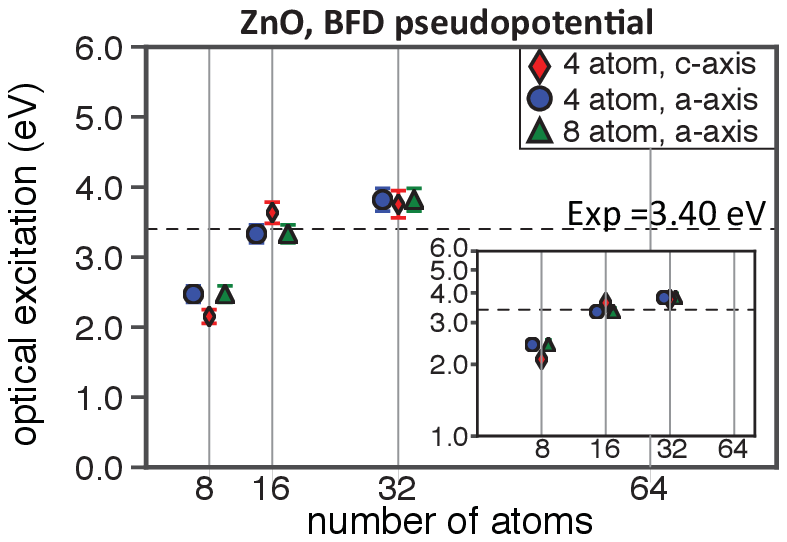} \\
\end{tabular} 
\caption {Optical excitation energy computed for different supercell sizes and shapes for ZnO with both (a) TN and (b) BFD pseudopotential. The excitation energy is obtained as the difference of the $\Gamma$ point ground state and first optically excited state energies. The optically excited state trial wave functions are constructed by removing an electron both from VBM orbitals aligned along the wurtzite $a$ and $c$ axes. } 
  \label{ZnOoptenergy}
\end{figure}

\begin{figure}[h]
\begin{tabular}{l} 
\includegraphics[width=7.5cm]{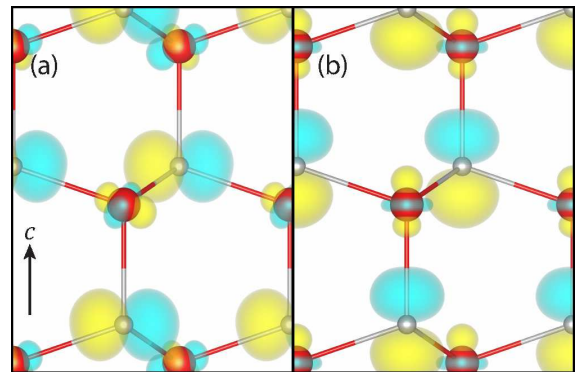} \\    
\end{tabular} 
\caption {Charge density of the DFT-PBE0 VBM orbitals at the $\Gamma$ point, using the TN pseudopotential. (a) orbital is aligned in the $a,b$ plane, and (b) orbital is parallel to the wurtzite $c$ axis.} 
  \label{ZnOValenceBand}
\end{figure}

Finally, we consider the calculation of the $\Gamma$-point optically excited state in ZnSe and ZnO. 
Since both of these semiconductors have a direct gap  at $\Gamma$, the calculated optical excitation energy will correspond to the optical band gap.  
Accurate calculation of band gaps will be a critical piece to establishing the functionality of the QMC method, so it is important to assess the scale and magnitude of finite size effects for such systems.  
To calculate the optical excitation energy, we have adopted a procedure that has been used successfully previously~\cite{ErtekinMgO}.  
The energy of the first optically excited state is calculated as 
\begin{equation}
OP = E_{\Gamma\rightarrow\Gamma}-E_g \hspace{1em}, 
\label{OP}
\end{equation} 
where $E_g$ is the ground state energy and $E_{\Gamma\rightarrow\Gamma}$ is the $\Gamma$-point optically excited state.  
The energy $E_{\Gamma\rightarrow\Gamma}$ is calculated by promoting an electron from the highest occupied Kohn-Sham orbital at $\Gamma$ to the lowest unoccupied orbital in the construction of the Slater determinant. 
Here, both terms $E_{\Gamma\rightarrow\Gamma}$, $E_g$ are evaluated only at the $\Gamma$ point. 
Since the optical excitation energy is an energy difference between two supercells of the same size and shape, the structure factor correction cancels out. 
Accordingly, one might expect that the dominant $1/N$ scaling exhibited for the ground state energies in Figs.~\ref{ZnSetotalenergy},\ref{ZnOtotalenergy} should not be present, leaving behind a faster convergence with increasing supercell size.
We are unaware of any existing detailed analysis of the scaling of finite size effects for optical excitation energies.
  
To analyze the behavior, in Figs.~\ref{ZnSeoptenergy},\ref{ZnOoptenergy} we report the computed gaps of ZnSe and ZnO using the same supercells that were considered in Figs. \ref{ZnSetotalenergy},\ref{ZnOtotalenergy}. 
Since the precise scaling of the finite size effect is unknown, Figs.~\ref{ZnSeoptenergy},\ref{ZnOoptenergy} show the computed gap as a function of supercell size on a linear/linear scale.
The insets show the same data plotted on a log/log scale to observe whether the scaling can be extracted. 
For ZnSe (Fig.~\ref{ZnSeoptenergy}), the $N=8$ atom supercells are too small, but it appears that all supercells with $N=16$ and $N=32$ are converged within error bars and our calculated optical gap of 2.8(2) eV is in agreement with experiment. From the inset there is no evidence of $1/N$ scaling, consistent with the expectation that the convergence is faster. 

The situation is more complicated for ZnO (Fig.~\ref{ZnOoptenergy}), which exhibits the wurtzite structure. 
It is symmetry-broken from cubic zincblende with unequal lattice constants in the $c$ and $a,b$ directions. 
Correspondingly, there are three orbitals at $\Gamma$ near the VBM which are slightly non-degenerate: one has orbitals aligned along the $c$ axis, and two have orbitals aligned within the $a,b$ plane.  
These are plotted in Fig.~\ref{ZnOValenceBand}. 
According to ARPES measurements, the symmetry breaking is small, $<0.1$ eV~\cite{arpes}. 

We address the BFD results for ZnO first, shown in Fig.~\ref{ZnOoptenergy}b.  
As for ZnSe the 8 atom supercells are too small, but the 16 and 32 atom supercells show more consistent trends, regardless of the supercell used. For the 32 atom supercells, our calculated excitation energy is 3.8(2) eV, which slightly overestimates the experimental value of 3.4 eV. 
In addition, the log/log plot in the inset also has no evidence of $1/N$ scaling, suggesting a faster convergence.  
We also tested the sensitivity of our calculated DMC optical excitation energies to the orbital from which the electron is removed in the calculation of the term $E_{\Gamma\rightarrow\Gamma}$ in Eq. (\ref{OP}). 
For the BFD pseudopotential, we find that (as expected) whether the electron is removed from the state with orbitals aligned parallel to the $c$ or the $a$ axis, our computed excitation energies are not sensitive.  
Overall, our calculated gaps for ZnO using the BFD pseudopotential overestimate the experimental value (3.4 eV) by a few tenths of an eV. 
These results are in line with self-consistent GW calculations, which find an GW gap of 3.84 eV \cite{PhysRevLett.96.226402}, although we should note that non self-consistent GW has a strong sensitivity to the calculation starting point for ZnO. 

The situation is different for the TN pseudopotential, however, for which our results are shown in Fig.~\ref{ZnOoptenergy}a.  
The optical excitation energy has been calculated using the same $N=8,16,32$ supercells that we used for the BFD pseudopotential.  
Compared to the BFD results, the trends in the computed TN optical excitations are less clear, and it appears that the finite size effects are not yet completely converged. 
Additionally there is an unexpected sensitivity of the excitation energy to the orbital from which the electron is removed in the calculation of $E_{\Gamma\rightarrow\Gamma}$. 
The reason for the different trends in Fig~\ref{ZnOoptenergy}a,b is not obvious, but the only difference between the two sets results is the pseudopotentials used for the calculations.
We conclude that the Burkatski-Fillipi-Dolg pseudopotential is better able to describe the optically excited state  wave function. 
Although there may be other differences as well, a key difference between the TN and the BFD pseudopotentials is the size of the core (Ar core vs. Ne core), so it is possible that the ionic cores may be more largely perturbed for the excited states. 
It is interesting that throughout this work the differences between the two pseudopotentials largely arose in the excited state simulations. 


\subsection{Conclusion}

In conclusion, we have carried out a detailed assessment of errors and uncertainties for the simulation of ZnSe and ZnO, arising from both controllable and uncontrollable approximations, within the fixed node diffusion Monte Carlo framework.  
We find that the both Trail-Needs and Burkatski-Filippi-Dolg Dirac-Fock pseudopotentials do an excellent job of the calculation of the atomization energy.  Localization errors introduced by non-local pseudopotentials can be very significant, particularly for pseudopotentials with large cutoff radii. 
For the ground state the magnitude of the fixed node error, as assessed by varying the nodes of the trial wave function, is comparatively small. 
Supercell finite size effects can have a significant effect on calculated ground state energies and extrapolation to the infinite supercell limit appears feasible by sampling supercells up to 16 or 32 atoms in size, particularly with the application of the $S(k)$ structure factor correction. 
When carried out systematically using the approach outlined here, the resulting atomization energies for ZnO and ZnSe are within 0.1 eV/atom of the experimental value, which is promising for the prediction of stability. 

For the calculation of optically excited states, assessment across a variety of supercell sizes and shapes is necessary. 
We observe that the computed excitations converge quickly with system size, with a scaling that is probably faster than linear.  
We find that the BFD pseudopotential is able to give a good description of the gap of ZnO. 
On the other hand, while the TN pseudopotential performed well for the gap of ZnSe, the results are not as clear for ZnO. 
These observations suggest that differences between pseudopotentials may be more prevalent excited state calculations even if not for the ground state, and that when transition metals are involved, small core pseudopotentials may be important for achieving high accuracy.

The detailed assessment presented here is expected to be of use to the QMC modeling community, towards the establishment of QMC ``best practices" for the simulations of solid materials.  
We also expect this to provide a foundation for further studies that exploit the accuracy available in QMC methods.

\begin{acknowledgements} 
We acknowledge financial support to J. Yu through the Computational Science and Engineering Fellowship Program at the University of Illinois.  L.K.W. was supported by the U.S. Department of Energy, Office of Science, Office of Advanced Scientific Computing Research, Scientific Discovery through Advanced Computing (SciDAC) program under Award Number FG02-12ER46875. This research is part of the Blue Waters sustained petascale computing project, which is supported by the National Science
Foundation (award No. OCI 07-25070 and ACI-1238993) and the state of Illinois. Blue Waters is a joint effort of the University of Illinois at Urbana--Champaign and its National
Center for Supercomputing Applications. Also, this research used resources of the Argonne Leadership Computing Facility, which is a DOE Office of Science User Facility supported under Contract DE-AC02-06CH11357.  
\end{acknowledgements}

%

\end{document}